\documentclass[12pt]{article}
\usepackage{graphicx}
\begin{document}
\rightline{NKU-08-SF1}
\bigskip
\begin{center}
{\Large\bf Quasinormal Modes   of   Charged Scalars  around  Dilaton Black Holes in 2 +1 Dimensions : Exact Frequencies}

\end{center}
\hspace{0.4cm}
\begin{center}
Sharmanthie Fernando \footnote{fernando@nku.edu}\\
{\small\it Department of Physics \& Geology}\\
{\small\it Northern Kentucky University}\\
{\small\it Highland Heights}\\
{\small\it Kentucky 41099}\\
{\small\it U.S.A.}\\

\end{center}

\begin{center}
{\bf Abstract}
\end{center}

\hspace{0.7cm} 

We have studied the charged scalar perturbation around a   dilaton black hole in 2 +1 dimensions. The wave equations of a massless charged scalar field is shown to be exactly solvable in terms of hypergeometric functions. The quasinormal frequencies are computed exactly. The relation between  the quasinormal frequencies and the charge of the black hole, charge of the scalar and the temperature of the black hole are analyzed. The asymptotic form of the real part of the quasinormal frequencies are evaluated exactly.


\section{Introduction}

When a black hole is perturbed by an external field, the dynamics of the 
scattered waves can be described in three stages \cite{frolov}. The first corresponds to the initial wave which will depend  on the source of disturbance. The second corresponds to the quasinormal modes with complex frequencies. Such modes are called quasinormal in contrast to normal modes since these are damped oscillations. The values of the quasinormal modes are independent of the initial disturbance and only depend on the parameters of the black hole. The focus of this paper is to analyze quasinormal modes  of a charge scalar around a dilaton black hole in 2 + dimensions. The last stage of perturbations is described by a power-law tail behavior of the corresponding field in some cases.

In recent times, there had been extensive work done to compute quasinormal modes (QNM) and to analyze them in various black hole backgrounds. A good review is  Kokkotas et. al.\cite{kokko}.

One of the reasons for the attention on QNM's is the conjecture relating anti-de-Sitter space (AdS) and conformal field theory(CFT) \cite{aha}. It is conjectured that  the imaginary part of the QNM's which gives the time scale to decay the black hole perturbations also corresponds to the time scale of the conformal field theory (CFT)  on the boundary to reach thermal equilibrium. There are  many works on AdS black holes on this subject \cite{horo} \cite{car1} \cite{moss} \cite{wang}. Also, if signals due to QNM's are detected by the gravitational wave detectors, one may be able to identify the charges of black holes and obtain  depper understanding of the inner structure of the black holes in nature. A recent review on QNM's and gravitational wave astronomy written by Ferrari and Gualtieri discuss such possibilities  \cite{ferr}.

There are many papers on the study of perturbations of black holes by neutral scalars. However, when a charged black hole is formed with gravitational collapse of charged matter, one expects perturbations by charged fields to develop out side the black hole. Hence, it is worthwhile to study charged scalar field perturbations. The late time evolution of a charged scalar in the gravitational collapse of charged matter to form Reissner-Nordstrom black holes were analyzed by Hod and Pirani \cite{hod1} \cite{hod2} \cite{hod3}. QNM's of massive charged scalar field around Reissner-Nordstrom black hole were studied by Konoplya \cite{kono1}. Decay of charged scalar and the Dirac field around Kerr-Newmann-de-Sitter black hole were studied by Konoplya and Zhidenko in \cite{kono2}. In \cite{kono3}, decay of massless charged scalars around variety of black holes in four dimensions were studied by Konoplya.

To the authors knowledge, most of  the work on QNM's of black holes in four and higher dimensions  are numerical except for few cases. Few we are aware of are, the massless topological black hole calculation done by Aros et. al. \cite{aros}, exact frequencies computed for gravitational perturbation of topological black holes in \cite{birs} and QNM computations for de Sitter space in \cite{ort1} \cite{ort2}. However, in 2+1 dimensions, QNM's can be computed exactly due to the nature of the wave equations. In particular,  the well known BTZ black hole \cite{banados} has been studied with exact results \cite{bir1} \cite{bir2} \cite{car2} \cite{abd}. The QNM's of the neutral scalars around the dilaton  black hole were computed exactly  in \cite{fer1}. The Dirac QNM's for the dilaton black hole was computed in \cite{ort3}. In this paper we take a step further by studying QNM's of a charged scalar around dilaton black holes in 2+1 dimensions which leads to exact results. To the authors knowledge, all work related QNM's of  charged scalars have been done numerically.

Extensions of the BTZ black hole with charge have lead to many interesting work. The first investigation was done by Banados et. al.\cite{banados}. Due to the logarithmic nature of the electromagnetic  potential , these solutions give rise to unphysical properties\cite{chan1}. The horizonless static solution with magnetic charge were studied by Hirshmann et.al.\cite{hirsh} and the persistence of these unphysical properties was highlighted by Chan \cite{chan1}. Kamata et.al.\cite{kamata} presented a rotating charged black hole with self (anti-self) duality imposed on the electromagnetic fields. The resulting solutions were asymptotic to an extreme BTZ black hole solution but had diverging mass and angular momentum \cite{chan1}. Clement \cite{clem}, Fernando and Mansouri\cite{fer3} introduced a Chern-Simons term as a regulator to screen the electromagnetic potential and obtained horizonless charged particle-like solutions. In this paper we consider an interesting class of black hole solutions obtained by Chan and Mann \cite{chan2}. The solutions represents static charged black holes with a dilaton field. It is a solution to low-energy string action. Furthermore, it has finite mass unlike some of the charged black holes described above.

We have organized the paper as follows: In section 2 an introduction to the geometry of the black hole is given. The charge scalar perturbation of the black hole is given  in section 3. The general solution to the wave equation is given in section 4. Solution with boundary conditions is given in section 5. QNM  frequencies  of the black hole is computed and analyzed in detail in section 6. Finally the conclusion is given in section 7.

\section{ Geometry of the static charged dilaton black hole}

In this section we will present the geometry and important details of the staic charged black hole. The Einstein-Maxwell-dilaton action which lead to these black holes considered by Chan and Mann \cite{chan2} is given as follows:
\begin{equation}
S = \int d^3x \sqrt{-g} \left[ R - 4  (\bigtriangledown \phi )^2 -
e^{-4  \phi} F_{\mu \nu} F^{\mu \nu} + 2 e^{4 \phi} \Lambda \right]
\end{equation}
Here, $ \Lambda$ is treated as the cosmological constant. In \cite{chan2}, it was discussed that black hole solutions exists only for $ \Lambda > 0$. Hence throught this paper we will treat $\Lambda > 0$.
The paramter $\phi$ is the dilaton field, $R$ is the scalar curvature and $F_{\mu \nu}$ is the Maxwell's field strength in the action. This action is conformally related to the low-energy string action in 2+1 dimensions. The static circularly symmetric solution to the above action is given by,

$$
ds^2= - f(r)dt^2 +  \frac{4 r^2 dr^2}{f(r)} + r^2 d \theta^2
$$
\begin{equation}
f(r) =\left( -2Mr + 8 \Lambda r^2 + 8 Q^2 \right); \hspace{0.1cm} \phi = \frac{1}{4}  ln (\frac{r}{\beta}) ; \hspace{1.0cm}F_{rt} = \frac{Q}{r^2}
\end{equation}
For $M \geq 8 Q \sqrt{\Lambda}$, the space-time represent a black hole. It has two horizons given by the zeros of $g_{tt}$;
\begin{equation}
r_+ =  \frac{M + \sqrt{ M^2 - 64 Q^2 \Lambda}}{8 \Lambda}; \hspace{1.0cm}
r_- = \frac{M - \sqrt{ M^2 - 64 Q^2 \Lambda}}{8 \Lambda}
\end{equation}
There is a singularity at $r=0$ and it is time-like. Note that in the presence of a non-trivial dilaton, the space-geometry of the black hole does not behave as either de-Sitter ( $\Lambda < 0$) or anti-de-Sitter ($\Lambda > 0$) \cite{chan2}. An important thermodynamical quantity corresponding to a black hole is the  Hawking temperature $T_H$. It is given by,
\begin{equation}
T_H= \frac{1}{4 \pi} |\frac{dg_{tt}}{dr}| \sqrt{-g^{tt} g^{rr}} |_{r=r_+} = \frac{M}{4 \pi r_+} \sqrt{ 1 - \frac{64 Q^2 \Lambda}{M^2}}
\end{equation}
The temperature $T_H=0$ for the extreme black hole with $M= 8 Q \sqrt{\Lambda}$. For the uncharged black hole $T_H = \frac{\Lambda}{ \pi}$. This black hole is also a solution to low energy string action  by a  conformal transformation,
\begin{equation}
g^{String} =  e^{4 \phi} g^{Einstein}
\end{equation}

In string theory,  it is possible to create charged solutions from uncharged ones  by duality transformations. For a review of such transformations see Horowitz\cite{horo1}. It is possible to apply such transformations to the uncharged blackhole with charge $Q =0$ in the metric in eq.(2) to obtain the charged blackhole with $Q \neq 0$. Such a duality was discussed in detail in the paper by Fernando \cite{fer1}.

\section{Charged scalar perturbation of  dilton  black holes}

We will develop the equations for a charged scalar field in the background of the static charged dilaton black hole in this section. The general equation for a massless charged scalar field in curved space-time can be written as,
\begin{equation}
\bigtriangledown ^{\mu} \bigtriangledown_{\mu}  \Phi  + (i e)^2 A^{\mu} A_{\mu} \Phi - 2 i e A^{\mu} \partial_{\mu} \Phi - i e \Phi \bigtriangledown^{\mu} A_{\mu}   =0
\end{equation}
Using the anzatz,
\begin{equation}
\Phi = e^{ i m \theta} \frac{\eta(t,r)} { \sqrt{r}}
\end{equation}
eq.(6)  simplifies into,
\begin{equation}
\frac{ \partial^2 \eta(t,r) }{ \partial t^2} - \frac{ \partial^2 \eta(t,r) }{ \partial r_{*}^2} + \frac{ 2 i e Q}{r}  \frac{\partial \eta(t,r)} { \partial t} + V(r) \eta(t,r)  =0
\end{equation}
Here, $V(r)$ is given by,
\begin{equation}
V(r) =  \frac{f(r) } { 2 r^{3/2} } \frac{d}{dr} \left( \frac{ f(r) } { 4 r^{3/2} } \right) + \frac{ m^2 f(r)}{r^2} - \frac{ e^2 Q^2 } { r^2}
\end{equation}
and $r_{*}$ is the tortoise coordinate computed as,
\begin{equation}
dr_{*} = \frac{2 r dr}{f(r)} \Rightarrow 
r_* = \frac{ 1}{ 4 \Lambda (r_+ - r_-)} \left( r_+ ln( (r - r_+) - r_- ln( r - r_-) \right)
\end{equation}
Note that when $r \rightarrow r_+$ , $r_* \rightarrow - \infty$ and for $r \rightarrow \infty$, $ r_* \rightarrow \infty $.
The function $f(r) $ is given by eq.(2) in section 2. By substituting the function $f(r)$ into the eq.(9), one can obtain a simplified version of the potential $V(r)$ as,
\begin{equation}
V(r) = -\frac{12 Q^4}{r^4} + \frac{ 4 M Q^2}{r^{3}} + \frac{1}{r^2} \left( - \frac{M^2}{4} + 8 m^2 Q^2 - 8 Q^2 \Lambda - e^2 Q^2 \right) - \frac{ 2 m^2 M}{ r} + ( 8 m^2 \Lambda + 4 \Lambda^2) 
\end{equation}
Note that if the function $\eta(t,r)$ is redefined as,
\begin{equation}
\eta(t,r) = e^{-i \omega t} \xi( r_{*} )
\end{equation}
the wave equation will simplifies to  the equation,
\begin{equation}
\left(\frac{d^2 }{dr_*^2} + \omega^2  + \frac{ 2 e Q \omega } { r}  - V(r) \right) \xi(r_*) = 0
\end{equation}
It is clear that if $e = 0$, eq.(13) becomes the  $Schrodinger$-type equation with a potential $V_{e=0}(r)$ given by,
\begin{equation}
V_{e=0}(r) = -\frac{12 Q^4}{r^4} + \frac{ 4 M Q^2}{r^{3}} + \frac{1}{r^2} \left( - \frac{M^2}{4} + 8 m^2 Q^2 - 8 Q^2 \Lambda  \right) - \frac{ 2 m^2 M}{ r} + ( 8 m^2 \Lambda + 4 \Lambda^2) 
\end{equation}
\begin{center}

\scalebox{0.9} {\includegraphics{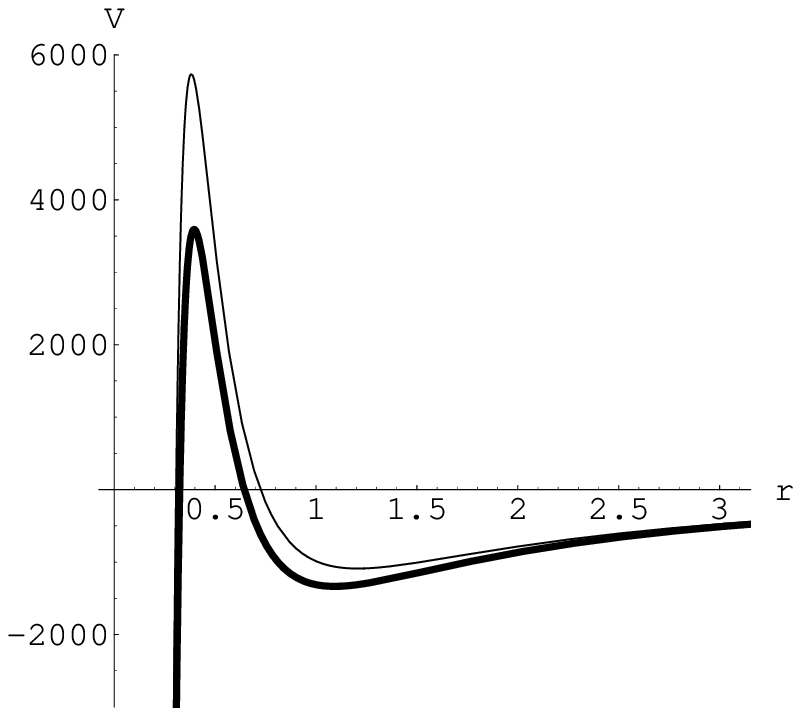}}

\vspace{0.3cm}
\end{center}
{Figure 1. The behavior of the potentials $V(r)$ and $V_{e=0}(r)$ with $r$ for $\Lambda=2$ $M=120$, $Q=3$, $m=2$ and $e = 6$. The dark curve represents $V(r)$ and the light curve represents $V_{e=0}(r)$}\\

The potentials are plotted in the Fig.1. Greater the value $e$ of the charged scalar, smaller the peak of the potential.

\section { General solution to the charged scalar wave equation}

In order to find exact solutions to the wave equation for the charged scalar, we will revisit the eq.(6) in section 3.
Using the anzatz,
\begin{equation}
\Phi =  e^{- i \omega t} e^{i m \theta} R(r) 
\end{equation}
eq.(6) leads to the radial equation,
\begin{equation}
\frac{d}{dr} \left( \frac{f(r)}{2} \frac{dR(r)}{dr} \right) + 2r^2 \left( \frac{\omega^2}{f(r)}   - \frac{m^2}{r^2}  \right)  R(r)  - \frac{ 4 e Q \omega r R(r)  } { f(r) }  + \frac{ 2 e^2 Q^2 R(r)  }{ f(r) }=0
\end{equation}
In order to solve the wave equation exactly, one can redefine  
$r$ coordinate of the eq.(16) with a new variable $z$ given by,
\begin{equation}
 z = \left( \frac{ r - r_+}{ r - r_-} \right)
\end{equation}
Note that in the new coordinate system, $z = 0$ corresponds to the horizon $r_+$ and $z = 1$ corresponds to infinity. With the new coordinate,  eq.(16) becomes,
\begin{equation}
z(1-z) \frac{d^2 R}{dz^2} + (1-z) \frac{d R}{dz} + P(z) R =0
\end{equation}
Here,
\begin{equation}
P(z) = \frac{A}{z} + \frac{B}{-1+z} + C
\end{equation}
where,
\begin{equation}
A=  \frac{(r_+ \omega  -e Q )^2}{ 16 (r_+- r_-)^2 \Lambda^2}; \hspace{1.0cm} B = \frac{ 8m^2 \Lambda - \omega^2} { 16 \Lambda^2}; \hspace{1.0cm} C = - \frac{(r_- \omega - e Q )^2}{16 (r_+- r_-)^2 \Lambda^2}
\end{equation}
Now, if   $R(z)$ is redefined as,
\begin{equation}
R(z) = z^{\alpha} (1-z)^{\beta} F(z)
\end{equation}
the radial equation given in eq.(18) becomes,
\begin{equation}
z(1-z) \frac{d^2 F}{dz^2} + \left(1 + 2 \alpha - (1+ 2 \alpha + 2\beta )z \right) \frac{d F}{dz} + \left(\frac{\bar{A}}{z} + \frac{\bar{B}}{-1+z} + \bar{C}     \right) F =0
\end{equation}
where,
$$\bar{A} = A + \alpha^2$$
$$ \bar{B} = B + \beta - \beta^2$$
\begin{equation}
 \bar{C} = C -(\alpha + \beta)^2 
\end{equation}
The above equation resembles the hypergeometric differential equation  which is of the form \cite{math},
\begin{equation}
z(1-z) \frac{d^2 F}{dz^2} + (c  - (1+a + b )z) \frac{d F}{dz} -ab  F =0
\end{equation}
By comparing the coefficients of eq.(22) and eq. (24), one can obtain the following identities,
\begin{equation}
c = 1+ 2 \alpha
\end{equation}
\begin{equation}
a+b = 2 \alpha + 2 \beta
\end{equation}
\begin{equation}
\bar{A}=A + \alpha^2 =0; \Rightarrow \alpha= \pm \frac{  i (r_+ \omega - e Q)}{ 4 \Lambda ( r_+ - r_-)}
\end{equation}
\begin{equation}
\bar{B} = B + \beta - \beta^2=0; \Rightarrow \beta = \frac{1 +  i \sqrt{ ( \frac{ \omega^2 - 8 m^2 \Lambda}{4 \Lambda} - 1 }) }{2}
\end{equation}
\begin{equation}
 ab = -\bar{C} = (\alpha + \beta)^2 - C 
\end{equation}
From eq.(26) and eq.(29),
$$a= \alpha + \beta + \gamma$$
\begin{equation}
b= \alpha + \beta -\gamma
\end{equation}
Here, 
\begin{equation}
\gamma= \sqrt{C} = \pm \frac{  i (r_- \omega - eQ )}{ 4 \Lambda ( r_+ - r_-)}
\end{equation}
With the above values for $a$, $b$, and $c$, The solution to the hypergeometric function $F(z)$ is given by \cite{math},
\begin{equation}
F(a,b,c;z) = \frac{\Gamma(c)} {\Gamma(a) \Gamma(b)} \Sigma \frac{ \Gamma(a+n) \Gamma( b+n)}{ \Gamma(c+n)}  \frac{z^n}{n!}
\end{equation}
with a radius of convergence being the unit circle $|z| =1$. Hence the general solution to the radial part of the charged scalar wave equation is given by,
\begin{equation}
R(z) = z^{\alpha} (1-z)^{\beta} F(a,b,c;z)
\end{equation}
with $a$, $b$, and $c$ given in the above equations. The general solution for the charged wave scalar equation is,
\begin{equation}
\Phi( z, t, \theta) = z^{\alpha} (1-z)^{\beta} F(a,b,c;z) e^{  i m \theta} e^ { - i \omega t}
\end{equation}

\section{Solution with boundary conditions}
In this section we will obtain solutions to the charged scalar with the boundary condition that the wave is purely ingoing at the horizon. The solutions are analyzed closer to the horizon and at infinity to obtain exact results for the wave function.

\subsection{ Solution at the near-horizon region}

First, the solution of the wave equation closer to the horizon is analyzed. For the charged black hole,
\begin{equation}
z = \frac{(r- r_+)}{ ( r - r_-)}
\end{equation} 
and as the radial coordinate $r$ approaches the horizon, $z$ approaches $0$. In the neighborhood of $z=0$, the hypergeometric function has two linearly independent solutions given by \cite{math}
\begin{equation}
F(a,b;c;z) \hspace{1.0cm} and \hspace{1.0cm} z^{(1-c)} F(a-c+1,b-c+1;2-c;z)
\end{equation}
Substituting the  values of $a,b,c$ in terms of $\alpha$, $\beta$, and $\gamma$, the general solution for $R(z)$ can be written as,
$$
R(z) = C_1 z^{\alpha} (1-z)^{\beta} F(\alpha + \beta + \gamma, \alpha+\beta - \gamma, 1+ 2 \alpha, z) 
$$
\begin{equation}
+ C_2 z^{-\alpha}(1-z)^{\beta} F( -\alpha + \beta + \gamma,-\alpha+\beta - \gamma,1-2 \alpha, z)
\end{equation}
Here, $C_1$ and $C_2$ are constants to be determined. Before proceeding any further, we want to  point out that the above equation is symmetric for $ \alpha \leftrightarrow - \alpha$. Note that in eq.(27), $\alpha$ could have both $\pm$ signs. Due to the above symmetry in eq.(37), we will choose the ``+" sign for $\alpha$  for the rest of the paper.

Since  closer to the horizon $z \rightarrow 0$, the above solution in eq.(37) approaches,
\begin{equation}
R(z  \rightarrow 0) = C_1 z^{\alpha}  + C_2 z^{-\alpha}
\end{equation}
Closer to the horizon, $r \rightarrow r_+$. Hence, $z$ can be approximated with
\begin{equation}
z \approx \frac{ r - r_+}{r_+ -  r_-}
\end{equation}
The ``tortoise'' coordinate for the charged black hole is given in eq.(10).
Near the horizon $r \rightarrow r_+$, the ``tortoise'' coordinate can be approximated to be 
\begin{equation}
r_* \approx  \frac{r_+}{4 \Lambda ( r_+ - r_-)} ln( r  - r_+)
\end{equation}
Hence,
\begin{equation}
r - r_+ = e^{ \frac{ 4 \Lambda ( r_+ - r_-)}{r_+} r_* }
\end{equation}
leading to,
\begin{equation}
z \approx \frac{ r - r_+}{ r_+ - r_-} = \frac{ 1 }{ (r_+ - r_-)} e^{ \frac{ 4 \Lambda ( r_+ - r_-)}{r_+} r_* }
\end{equation}
Hence eq.(38) can be re-written in terms of $r_*$ as,
\begin{equation}
R(r \rightarrow r_+) =  C_1 \left(\frac{ 1} { r_+ - r_-} \right)^{\alpha} e ^{ i  \hat{\omega} r_*}  + C_2 \left(\frac{1}{ r_+ - r_-} \right)^{ - \alpha} e^{ -i  \hat{\omega} r_*}
\end{equation}
To obtain the above expression, $\alpha$  is substituted from eq.(27) and
\begin{equation}
\hat{\omega} = \omega - \frac{ e Q} { r_+}
\end{equation}
The first and the second term in eq.(43) corresponds to the outgoing and the ingoing wave respectively. Now, one can impose the condition  that  the wave is purely ingoing at the horizon. Hence we pick $C_1 = 0$ and $C_2 \neq 0$. Therefore the solution closer to the horizon is,
\begin{equation}
R(z \rightarrow 0 ) =  C_2 z^{-\alpha} (1-z)^{\beta} F(-\alpha + \beta + \gamma, -\alpha+\beta - \gamma, 1 - 2 \alpha, z) 
\end{equation}

\subsection{Solution at asymptotic region}

Now the question is what the wave equation is when $r \rightarrow \infty$. For large $r$, the function $f(r) \rightarrow 8 \Lambda r^2$. When $f(r)$ is replaced with this approximated function in the wave equation given by eq.(16), it simplifies to,

\begin{equation}
\frac{d}{dr} \left( 4 \Lambda r^2 \frac{dR(r)}{dr} \right) + 2r^2 \left( \frac{\omega^2}{ 8 \Lambda r^2}   - \frac{m^2}{r^2}  \right)  R(r)  - \frac{  e Q \omega  R(r)  } { 2 \Lambda r  }  + \frac{  e^2 Q^2 R(r)  }{ 4 \Lambda r^2}=0
\end{equation}
For large $r$, one can neglect the last two terms in the above equation. Hence finally, the wave equation at large $r$ can be  expanded to be,
\begin{equation}
r^2 R'' + 2 r R' + p R =0
\end{equation}
where, 
\begin{equation}
p = \frac{\omega^2} { 16 \Lambda^2} - \frac{m^2}{2 \Lambda}
\end{equation}
One can observe that $ p = -B$ from eq.(20). Also eq.(47) is the well known Euler equation with the solution,
\begin{equation}
R(r) = D_1 \left( \frac{r_+ - r_-}{r} \right)^{a_1} + D_2 \left( \frac{r_+ - r_- }{r} \right)^{a_2}
\end{equation}
with,
\begin{equation}
a_1=   \frac{ 1 + \sqrt{ 1 - 4 p} }{2} =  \beta; \hspace{1.0 cm}
a_2 =  \frac{ 1 - \sqrt{ 1 - 4 p} }{2} = (1- \beta)
\end{equation}
The expression for $\beta$ is given in eq.(28). Note that the form in eq.(49) is chosen to facilitate to compare it with the matching solutions in section 5.3.

\subsection{Matching the solutions at the near horizon and the asymptotic region}

In this section we  match the asymptotic solution given in eq.(49) to the large $r$ limit (or the $z \rightarrow 1$ ) of the near-horizon solution given in 
eq.(45) to obtain an exact expression for $D_1$ and $D_2$.  To obtain the $z \rightarrow 1$ behavior of eq. (45), one can perform a well known transformation on hypergeometric function given as follows \cite{math}
$$
F(a,b,c,z) = \frac{ \Gamma(c) \Gamma(c-a-b)}{\Gamma(c-a) \Gamma(c-b)} F(a,b;a+b-c+1;1-z) 
$$
\begin{equation}
 +(1-z)^{c-a-b}\frac{ \Gamma(c) \Gamma(a+b-c)}{\Gamma(a) \Gamma(b)} F(c-a,c-b;c-a-b+1;1-z)
\end{equation}
Applying this transformation to eq.(45) and substituting for the values of $a,b,c$, one can obtain the solution to the wave equation in the asymptotic region as follows;
$$
R(z) = C_2 z^{-\alpha} (1-z)^{\beta} \frac{ \Gamma(1 - 2 \alpha) \Gamma(1 - 2 \beta)}{\Gamma(1 - \alpha - \beta - \gamma) \Gamma( 1 - \alpha - \beta + \gamma)} F( -\alpha + \beta + \gamma, -\alpha + \beta -\gamma; 2 \beta ;1-z) $$
\begin{equation}
+ C_2  z^{-\alpha} (1-z)^{1 - \beta} \frac{ \Gamma( 1 - 2 \alpha ) \Gamma( -1 + 2 \beta )}{\Gamma( -\alpha + \beta + \gamma) \Gamma( -\alpha + \beta - \gamma)} F( 1 - \alpha - \beta -\gamma, 1 - \alpha - \beta + \gamma ; 2 - 2 \beta;1-z)
\end{equation}
Now we can take the limit of  $R(z)$ as $ z \rightarrow 1$ ( or $r \rightarrow \infty$) which will lead to,
$$
R(z \rightarrow 1) = C_2  (1-z)^{\beta} \frac{ \Gamma(1 - 2 \alpha) \Gamma(1 - 2 \beta)}{\Gamma(1 - \alpha - \beta - \gamma) \Gamma( 1 - \alpha - \beta + \gamma)} 
$$
\begin{equation}
+ C_2 (1-z)^{1 - \beta} \frac{ \Gamma( 1 - 2 \alpha ) \Gamma( -1 + 2 \beta )}{\Gamma( -\alpha + \beta + \gamma) \Gamma( -\alpha + \beta - \gamma) } \end{equation}
Note that we have replaced $F(a,b,c,1- z)$ and $z^{\alpha}$ with 1  when $z$ approaches 1.
Since,
\begin{equation}
1 - z = \frac{r_+ - r_-}{r- r_-},
\end{equation}
for large $r$, the above can be approximated with,
\begin{equation}
1 - z \approx \frac{r_+ - r_-}{r }
\end{equation}
By replacing $1 - z$ with the above expression in eq.(55), $R(r)$ for large $r$ can be written as,
$$
R(r \rightarrow \infty) = C_2  \left(\frac{r_+ - r_-}{r }\right)^{\beta} \frac{ \Gamma(1 - 2 \alpha) \Gamma(1 - 2 \beta)}{\Gamma(1 - \alpha - \beta - \gamma) \Gamma( 1 - \alpha - \beta + \gamma}
$$
\begin{equation} 
+ C_2  \left(\frac{r_+ - r_-}{r }\right)^{1 - \beta} \frac{ \Gamma( 1 - 2 \alpha ) \Gamma( -1 + 2 \beta )}{\Gamma( -\alpha + \beta + \gamma) \Gamma( -\alpha + \beta - \gamma) } 
\end{equation}
By comparing eq.(49) and eq.(56), the coefficients $D_1$ and $D_2$ can be written as,
\begin{equation}
D_1 = C_2 \frac{ \Gamma(1 - 2 \alpha) \Gamma(1 - 2 \beta)}{\Gamma(1 - \alpha - \beta + \gamma) \Gamma( 1 - \alpha - \beta - \gamma) } 
\end{equation}
\begin{equation}
D_2 = C_2  \frac{ \Gamma( 1 - 2 \alpha ) \Gamma( -1 + 2 \beta )}{\Gamma( -\alpha + \beta + \gamma) \Gamma( -\alpha + \beta - \gamma )} 
\end{equation}
To determine which part of the solution in eq.(49) corresponds to the ``ingoing'' and ``outgoing'' respectively, we will first find the tortoise coordinate $r_{*}$ in terms of $r$ at large r. Note that for large $r$, $f(r) \rightarrow 8 \Lambda r^2 $. Hence  the equation relating the tortoise coordinate $r_*$ and $r$ in eq.(10) simplifies to,
\begin{equation}
dr_{*} = \frac{ dr}{ 4 \Lambda r}
\end{equation}
The above can be integrated to obtain,
\begin{equation}
r_{*} \approx  \frac{1}{4 \Lambda} ln( \frac{ r}{r_+} )
\end{equation}
Hence,
\begin{equation}
r \approx r_+ e^{ 4 \Lambda r_*}
\end{equation}
Substituting $r$ from eq.(61) and $\beta$ from eq.(28) into the eq.(49), $R(r \rightarrow \infty)$ is rewritten as,
$$
R(r  \rightarrow \infty ) \rightarrow D_1 \left( \frac{r_+ - r_-}{r_+}\right)^{\beta}  e ^{ -i  \omega r_* \sqrt{1 - \frac{ 4 \Lambda^2}{ \omega^2} ( \frac{2 m^2}{\Lambda} +1) } - 2 \Lambda r_*} $$
\begin{equation}
+ D_2 \left( \frac{r_+ - r_-}{r_+}\right)^{1 - \beta}  e ^{ i  \omega r_* \sqrt{1 - \frac{ 4 \Lambda^2}{ \omega^2} ( \frac{2 m^2}{\Lambda} +1) } - 2 \Lambda r_*} 
\end{equation}
From the above it is clear that the first term and the second term represents the ingoing and outgoing waves respectively.

\section{Quasinormal modes of the dilaton black hole}

Quasi normal modes of a classical perturbation of black hole space-times are defined as the solutions to the related wave equations with purely ingoing waves at the horizon. In addition, one has to impose boundary conditions on the solutions at the asymptotic region as well. In asymptotically flat space-times, the second boundary condition is the solution to be purely outgoing  at spatial infinity. For non-asymptotically flat space times, there are two possible boundary conditions to impose at sufficiently large distances from the black hole horizon: one, is the field to vanish at large distances and the other is for the flux of the field to vanish at far from the horizon. Here, we will choose the first. This is the condition imposed in reference \cite{fer1}. Another example in 2+1 dimensions where the vanishing of the filed at large distance is imposed is given in reference \cite{bir1} where QNM's of scalar perturbations of BTZ black holes were computed exactly.

Let's consider the field $R(r)$ at large distances given by eq.(56). Clearly the second term vanishes when $ r \rightarrow \infty$. This  also can be seen from eq.(52) where the second term vanishes for $ z \rightarrow 1$. Since $C_2$ is not zero, the first term vanish only at the poles of the Gamma functions $\Gamma(1 - \alpha - \beta + \gamma)$ or $\Gamma( 1 - \alpha - \beta - \gamma)$. Note that the Gamma function $\Gamma(x)$ has poles at $ x = - n$ for $ n = 0,1,2..$.Hence to obtain QNM's, the following relations has to hold.
\begin{equation}
1- \alpha - \beta - \gamma = - n
\end{equation}
or 
\begin{equation}
1 - \alpha - \beta + \gamma = -n
\end{equation}
The above two equations leads to two possibilities for $\beta$ as follows,
\begin{equation}
\beta = ( 1 + n) - \alpha \pm \gamma
\end{equation}
We want to recall here that $\gamma$ in eq.(31) could have both  signs. Due to the nature of eq.(65), there is no need to choose a specific sign to proceed from here. The two possibilities leads to two equations for $\beta$ given by,
\begin{equation}
\beta = ( 1 + n) - \frac{ i \omega } { 4 \Lambda }
\end{equation}
and
\begin{equation}
\beta = ( 1 + n) - i ( \kappa_1 \omega - \kappa_2 e Q )
\end{equation}
where,
\begin{equation}
\kappa_1 = \frac{ 1 }{ 4 \Lambda} \left( \frac{ r_+ + r_-}{ r_+ - r_-} \right); \hspace{1 cm} \kappa_2 = \frac{ 1} { 2 \Lambda ( r_+ - r_- ) }
\end{equation}
 By combining the  above equations with the eq.(28) given by,
\begin{equation}
\frac{m^2}{  2 \Lambda} - \frac{ \omega^2}{ 16 \Lambda^2} = - \beta + \beta^2
\end{equation}
one can obtain the quadratic equation for $\omega$ given by,
\begin{equation}
\omega^2 \left( \frac{1}{  ( 16 \Lambda^2 \kappa_1^2 -  1)} \right) + \omega  \left( i( 2n + 1 ) \kappa_1 - 2 \kappa_1 \kappa_2 e Q \right) + \left( \frac{m^2}{ 2 \Lambda} - n^2 - n - i ( 2n +1) \kappa_1 e Q + (\kappa_2 e Q )^2 \right)
\end{equation}
Note that the $\beta$ in eq.(66) corresponds to the QNM's of the neutral scalars for the uncharged black hole with  $Q=0$ leading to  $r_- = 0$. Hence by taking $r_-= 0$ in $\kappa_1$, one recover the quadratic equation for the neutral scalar for  $ Q = 0 $.
One can solve th above quadratic equation to obtain exact values of QNM frequencies $\omega$. There are three cases on can consider: QNM's of neutral scalars ( for $Q = 0$ and $ Q \neq 0$) and charged scalars. The QNM's of the neutral scalars were analyzed in detail in the paper by Fernando \cite{fer1}. We will any way state the results in order to compare the QNM's of the charged scalars in the following section.

\subsection{QNM frequencies  of neutral scalars with $e =  0$ }

By letting $e=0$ in the quadratic equation given above, one can solve it for $\omega$ as discussed in \cite{fer1}. First  one can consider the QNM's for the uncharged black hole with $Q =0$. The solution for $\omega$ is  given  as,
\begin{equation}
\omega =  \frac{-2 i}{ 2n +1} \left( 2 \Lambda n (1+n) - m^2 \right)
\end{equation}
They are pure imaginary. Due to the minus sign in front, these oscillations will be damped leading to stable perturbations for $2 \Lambda n (1+n) > m^2$. However, for $2 \Lambda n (1+n) < m^2$, the oscillations would lead to unstable modes. This was pointed out in \cite{ort3}.

One can also compute the QNM's for the neutral scalar for the charged dilaton black hole with $Q \neq 0$ as,
\begin{equation}
\omega=   \frac{-i}{  ( 16 \Lambda^2 \kappa_1^2 -  1)} \left(   8 \Lambda^2 \kappa_1(1+2 n) + 2  \sqrt{ 2m^2 \Lambda ( 16 \Lambda^2 \kappa_1^2 - 1) + 4 \Lambda^2 (4 \Lambda^2 \kappa_1^2  +  n^2 + n )}   \right)
\end{equation}
Note that $16 \Lambda^2 \kappa_1 >1$ and $\omega$ will always be pure imaginary. Also, due to the minus sign in front, these oscillations will be damped leading to stable neutral scalar perturbations.

\subsection{QNM's of the charged scalar with $ e \neq 0$}
Now, one can solve the eq.(70) to obtain the exat results for the QNM frequencies for the charged scalar as,

$$
\omega=   \frac{1}{  ( 16 \Lambda^2 \kappa_1^2 -  1)} \left(  -i 8 \Lambda^2 \kappa_1(1+2 n) + 16 e \kappa_1 \kappa_2 Q \Lambda^2 - \right.
$$
\begin{equation}
\left. 2 i \sqrt{ 2m^2 \Lambda ( 16 \Lambda^2 \kappa_1^2 - 1) + 4 \Lambda^2 (4 \Lambda^2 \kappa_1^2  +  n^2 + n ) - 4 e^2 \kappa_2^2 Q^2 \Lambda^2 + 4 i e \kappa_2 Q \Lambda^2 ( 2 n + 1)} \right)
\end{equation}
$\omega$ is not pure imaginary in this case: it has a real part which depends on $e$. For $e \rightarrow 0$, the above QNM approaches  to the values for the neutral scalar in eq.(72). To separate the real part and the imaginary part of $\omega$, the part inside the square root is redefined as follows;
Let the parameters $z_1$, $z_2$, $\rho$ and $Z$ defined as,
\begin{equation}
z_1 = 2m^2 \Lambda ( 16 \Lambda^2 \kappa_1^2 - 1) + 4 \Lambda^2 (4 \Lambda^2 \kappa_1^2  +  n^2 + n ) - 4 e^2 \kappa_2^2 Q^2 \Lambda^2
\end{equation}
\begin{equation}
z_2 = 4  e \kappa_2 Q \Lambda^2 ( 2 n + 1)
\end{equation}
\begin{equation}
Z = \sqrt{ z_1^2 + z_2^2 }
\end{equation}
\begin{equation}
\rho = tan^{-1} \left( \frac{z_2}{z_1} \right)
\end{equation}
Then, $\omega = \omega_{real} + i \omega_{imaginary}$ can be separated with,

\begin{equation}
\omega_{real} = \frac{1}{  ( 16 \Lambda^2 \kappa_1^2 -  1)} \left(16 e \kappa_1 \kappa_2 Q \Lambda^2 + 2 \sqrt{ Z} Sin( \rho/2) \right)
\end{equation}
\begin{equation}
\omega_{imaginary} = \frac{1}{  ( 16 \Lambda^2 \kappa_1^2 -  1)} \left(  -8 \Lambda^2 \kappa_1(1+2 n) - 2 \sqrt{Z} Cos( \rho/2) \right)
\end{equation}
When $ e \rightarrow 0$, $ \rho \rightarrow 0$ which leads to $\omega_{real} \rightarrow 0$ as expected. 
\\

In the Figure. 2, $\omega_{imaginary}$ is plotted against the charge $Q$ of the black hole. It is clear that the magnitude of $\omega_{imaginary}$ is larger for charged scalar in comparison with the neutral scalar. Hence, the neutral scalar decays slower compared to the charged scalar. Similar behavior was observed in the charged scalar decay compared to the neutral scalar in Reissner-Nordstrom and Reissner-Nordstrom anti-de-Sitter black hole \cite{kono3}.
\begin{center}

\scalebox{0.9} {\includegraphics{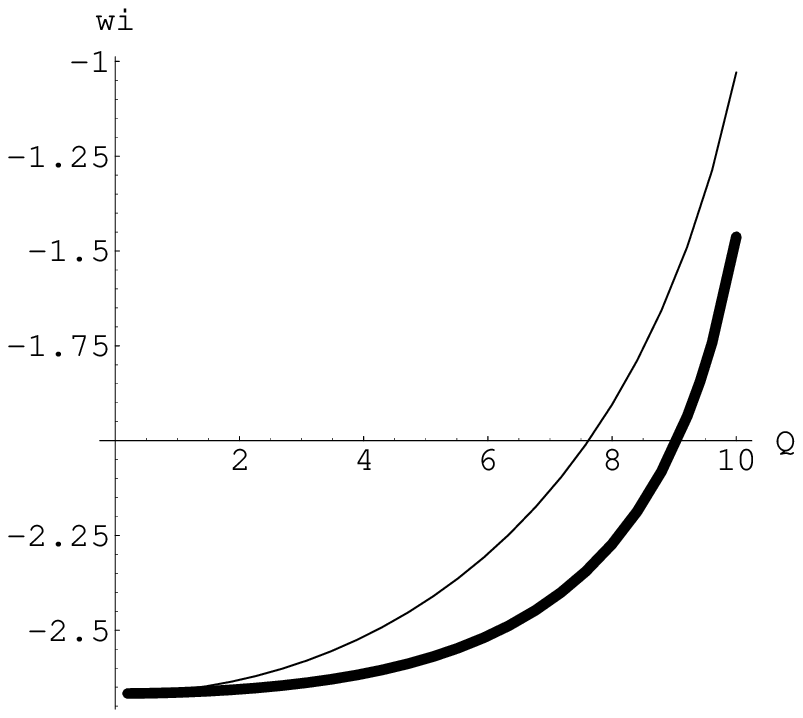}}

\vspace{0.3cm}
\end{center}

{Figure 2. The imaginary part  of $\omega$ Vs $Q$ for $\Lambda=2$ $M=120$, $m=2$ and $n = 1$. The dark curve represents the curve for  $e = 4$ and the light curve represents for  $ e = 0$}\\

Next, we observe the behavior of $\omega$ Vs charge $e$ for two different values of black hole charge $Q$ as given in the Figure.3. Higher the $Q$, larger the $\omega_{imaginary}$. Similarly, the real part of $\omega$ is larger for large $Q$ as given in Figure.4.

\begin{center}
\scalebox{0.9} {\includegraphics{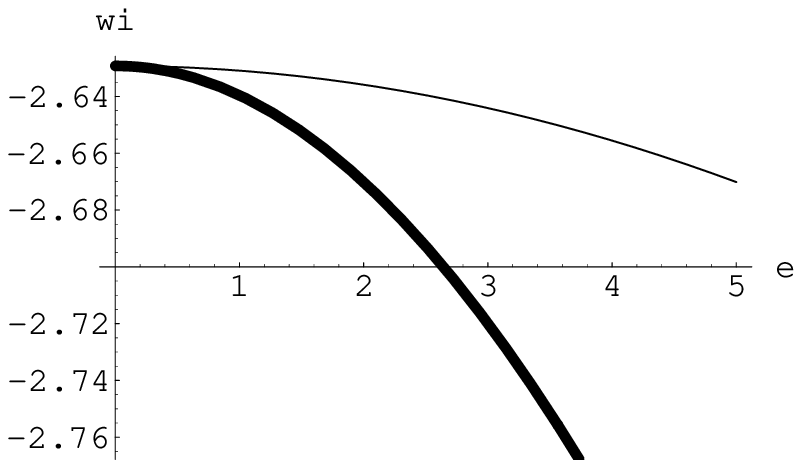}}

\vspace{0.3cm}
\end{center}
{Figure 3. The imaginary part of $\omega$ Vs $e$ for $\Lambda=2$ $M=120$, $m=2$ and $n = 1$. The dark curve represents the curve for  $Q = 5$ and the light curve represents for  $ Q = 2$}

\begin{center}

\scalebox{0.9} {\includegraphics{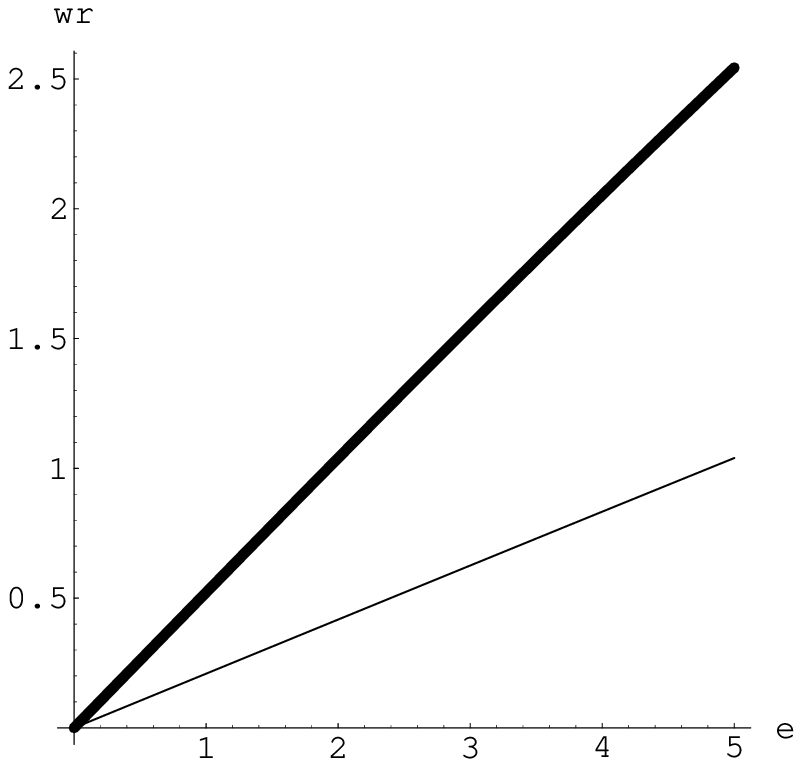}}

\vspace{0.3cm}

\end{center}
{Figure 4. The real part of $\omega$ Vs $e$ for $\Lambda=2$ $M=120$, $m=2$ and $n = 1$. The dark curve represents the curve for  $Q = 5$ and the light curve represents for  $ Q = 2$}\\

In Figure. 5, $\omega_{imaginary}$ is plotted Vs the temperature of the black hole. For both the neutral scalar and the charged scalar, there is a linear behavior of $\omega_{imaginary}$ Vs T.

\begin{center}

\scalebox{0.9} {\includegraphics{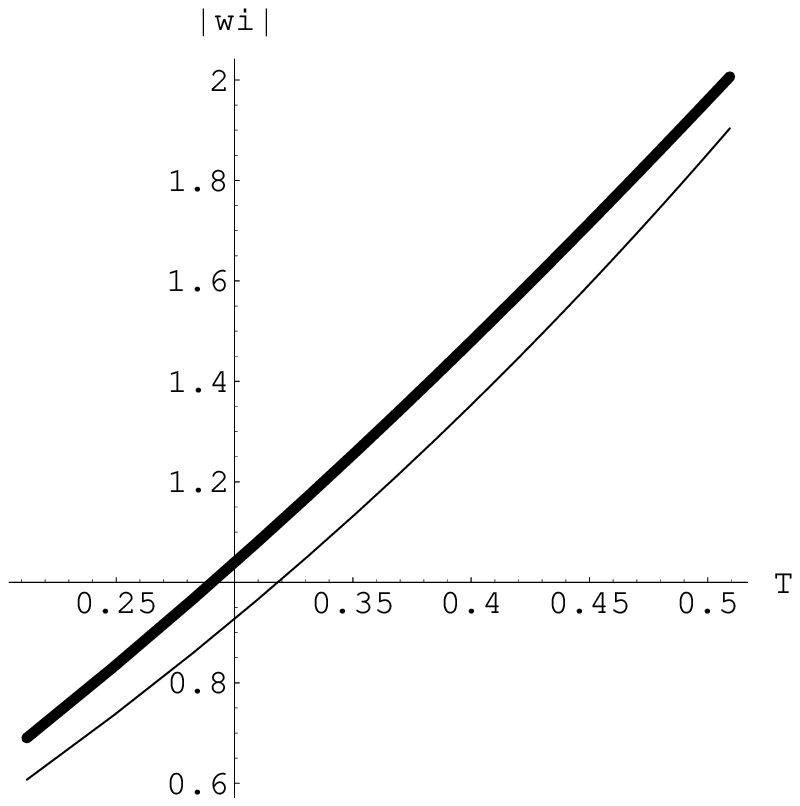}}

\vspace{0.3cm}

\end{center}
{Figure 5. The imaginary part  of $\omega$ Vs $T$ for $\Lambda=2$ $r_{-}=2$, $m=2$, and $n = 1$. The dark curve represents the curve for fixed $e = 2$ and the light curve represents for fixed $ e = 0$}\\

In Figure. 6, the behavior of $\omega_{imaginary}$ is plotted Vs the horizon radius $r_+$. It is concluded that  for the same $r_+$, the neutral scalar has a smaller decay rate than the charged scalar.
\begin{center}

\scalebox{0.9} {\includegraphics{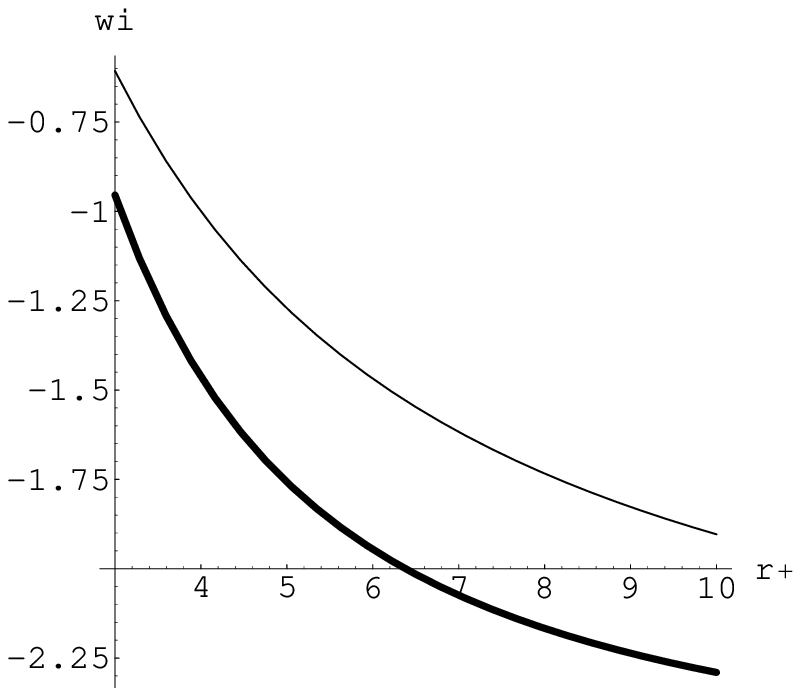}}

\vspace{0.3cm}

\end{center}
{Figure 6. The imaginary part  of $\omega$ Vs $r_+$ for $\Lambda=2$ $r_m = 2$, $m=2$ and $n = 1$. The dark curve represents the curve for  $e = 4$ and the light curve represents for fixed $ e = 0$}\\

As noted  in the introduction, there are several papers focused on computing the asymptotic value of the $\omega_{real}$ of black holes with regard to the quantization of the black holes. In the Figure. 7, $\omega_{real}$ is plotted Vs $n$. It is observed that it reaches a constant for large $n$. The asymptotic from of the real part of QNM is computed taking the limit of $\omega_{real}$ as $ n \rightarrow \infty$: the value is simply,
\begin{equation}
\omega_{real} ( n \rightarrow \infty ) = \frac{ e \sqrt{ r_p \Lambda} }{ \sqrt{r_m}}
\end{equation}

\begin{center}

\scalebox{0.9} {\includegraphics{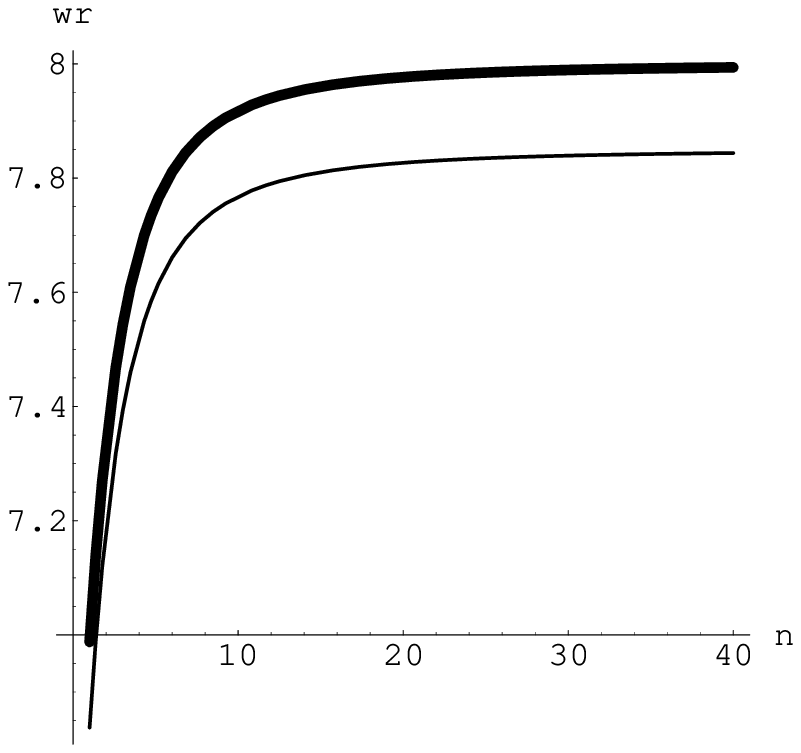}}

\vspace{0.3cm}

\end{center}
{Figure 7. The real part  of $\omega$ Vs $n$ for $\Lambda=2$, $r_p= 4$, $ r_m = 2$ and  $m=2$. The dark curve represents the curve for  $e = 4$ and the light curve represents for fixed $ e = 3.9$}\\

\section{Conclusion}

We have studied the perturbation of the dilaton black hole in 2+1 by a charged scalar. The wave equations are solved exactly as  hypergeometric functions. The QNM frequencies  are computed exactly. It is observed that the QNM's have both a real and an imaginary component. The QNM's of the  neutral scalars were pure imaginary \cite{fer1}. Also it is noted that the charged scalars decay faster compared to the neutral scalars for a given black hole. This observation is in agreement with the behavior observed by Konoplya \cite{kono3} in Reissner-Nordstrom and Reissner-Nordstrom-anti-de-Sitter black holes in four dimensions. The behavior of $\omega$ with various parameters are analyzed in detail. We observe the linear relation of $\omega_{imaginary}$ with the temperature of the black hole. Similar observations were reported for QNM frequencies of higher dimensions in AdS space in \cite{horo}. The asymptotic value of $\omega_{real}$ is computed to be $\frac{ e \sqrt{ r_p \Lambda} }{ \sqrt{r_m}}$.

It would be interesting to compute the greybody factors and particle emission rates for the charged scalars for this black hole. The greybody factors were studied for the neutral scalar in  \cite{fer2}. Since the wave equation is been already solved, it should be a welcome step towards  understanding the Hawking radiation from these black holes. There are few works related to such computations of charged particles: the particle emission by charged leptons from non rotating black holes by Page \cite{page} and emission of charged particles by four and five dimensional blackholes by Gubser and Klebanov \cite{gub}.

Since the asymptotic values of the real part of the QNM frequencies are computed exactly, it would be interesting to study the area spectrum of these black holes along the lines of the work by Setare \cite{set1} \cite{set2}.

Another interesting avenue to proceed would be to analyze the QNM's of the extreme dilaton black hole studied in this paper. Some  extreme black holes  have proven to be supersymmetric. For example, the extreme Reissner-Nordstrom black hole is shown to be supersymmetric since it can be embedded in N=2 supergravity theory \cite{gib}. Onozawa et.al \cite{ono} showed that the QNM's of the extreme Reissner-Nordstrom blackhole for spin 1, 3/2 and 2 are the same. If it is possible to find a suitable supergravity theory to embed the  dilaton black hole in this paper, one may be able to observe if extremality plays a role in it. Hence it would be interesting to compute the QNM's for the extreme dilaton black hole in 2 + 1 dimensions for the charged Dirac fields and vector fields along with the charged scalar to understand such behavior in low dimensions.

The dilaton black hole considered in this paper is one of the most favorable charged black holes in 2 +1 dimensions to study many  issues discussed above in a simpler setting with exact values for QNM frequencies.

\end{document}